\documentstyle[twocolumn]{article}
\title{Comment on ``Experimental Nonlocality Proof of Quantum
Teleportation and Entanglement Swapping''} 
\author{
Eric Dennis\thanks{\tt edennis@princeton.edu}\\
Chemistry Dept.\\
Princeton University\\
Princeton, NJ 08544}

\begin{document}

\maketitle

In a recent letter, Jennewein \emph{et al.} \cite{paper} reported an
experiment demonstrating entanglement swapping among two pairs (0,1 and
2,3) of down-converted photons. They accomplished this by performing a
Bell measurement on photons 1 and 2, and they verified the effect with
regular polarization measurements on photons 0 and 3. By selecting data
from runs in which 1,2 are measured to be in the Bell state
$|\psi^-\rangle$, they find that the measurement results for 0,3 violate a
Bell (CHSH) inequality, and they conclude that 0,3 have exhibited quantum
nonlocality.

A peculiarity is noted for the case in which the Bell measurement on 1,2
is performed only \emph{after} 0,3 have their polarizations measured.  In
this case, Jennewein \emph{et al.} say that the measurements on 0 and 3
``indicate... that photons 0 and 3 were entangled.''

But at the time 0 and 3 were measured, they were not entangled, because
they started out unentangled and no entangling operation had been
performed up to that point. They cannot be entangled retroactively. We
are thus lead to conclude that photons 0 and 3 have violated a Bell
inequality while not even being entangled with each other!

In fact, while the data for 0 and 3 does technically violate a Bell
inequality in this case, it is due not to quantum nonlocality between 0
and 3, but rather to the post-selection of 0,3 data according to whether
or not 1,2 are measured to be in the Bell state $|\psi^-\rangle$. 

As far as Bell inequalities are concerned, this kind of post-selection is
the same as enhancing observed correlations by manually discarding
selected runs after comparing the 0,3 polarization records. In these
experiments, the measurement results for photons 1 and 2 have simply been
used as markers for such runs. Indeed 0 was entangled with 1, and 2 with
3, but this entanglement has nothing to do with the Bell inequality
violations involving 0 and 3.

On the other hand, in the experiments where 1 and 2 are measured first, 0
and 3 are in fact projected into an entangled state and the resulting Bell
inequality violation does indicate real quantum nonlocality.

\end{document}